\documentclass[preprint,12pt]{elsarticle}

\usepackage{graphicx}
\usepackage{amssymb}
\usepackage{amsthm}
\usepackage{epstopdf}
\usepackage{amsfonts}
\usepackage{euscript}

\usepackage{multirow}
\usepackage{array}
\usepackage{amsmath}

\usepackage{booktabs}
\usepackage{diagbox,tabularx,blindtext}

\journal{Journal Name}

\begin{document}

\begin{frontmatter}


\title{Mutual transitions between stationary and moving dissipative solitons}



\author{Bogdan Kochetov}
\address{State Key Laboratory of Integrated Optoelectronics, \\
College of Electronic Science and Engineering, International Center of Future Science,\\ Jilin University, 2699 Qianjin St., Changchun 130012, China}
\ead{kochetov@jlu.edu.cn}

\begin{abstract}
Application of stable localized dissipative solitons as basic carriers of information promises the significant progress in the development of new optical communication networks. The success development of such systems requires getting the full control over soliton waveforms. In this paper we use the fundamental model of dissipative solitons in the form of the complex Ginzburg-Landau equation with a potential term to demonstrate controllable transitions between different types of coexisted waveforms of stationary and moving dissipative solitons. Namely, we consider mutual transitions between so-called plain (fundamental soliton), composite, and moving pulses. We found necessary features of transverse spatial distributions of locally applied (along the propagation distance) attractive potentials to perform those waveform transitions. We revealed that a one-peaked symmetric potential transits the input pulses to the plain pulse, while a two-peaked symmetric (asymmetric) potential performs the transitions of the input pulses to the composite (moving) pulse.
\end{abstract}


\end{frontmatter}


\section{Introduction}
\label{S:1}
Dissipative solitons are known as stable spatially localized structures existing in extended nonlinear dissipative systems far from equilibrium \cite{Akhmediev_Book1,Akhmediev_Book2,Purwins_AP_2010,Liehr_Book}. They were found in different open nonlinear systems having the applications ranged from hydrodynamics, optics, and condensed matter to chemistry and biology. On the other hand, the development of dissipative solitons is a form of the self-organization linking the animate and the inanimate. These circumstances have continuously stimulated both experimental search of unusual features of dissipative solitons in particular systems and development of their fundamental mathematical models over the last few decades. As a result of the latter activity, basic models describing the evolution of dissipative solitons have been found in the form of the complex Ginzburg-Landau equation (CGLE) \cite{Cross_RMP_1993, Aranson_RMP_2002, García-Morales_CP_2012}. The CGLE accounts for such important features of dissipative systems with inertialess nonlinearity as loss, gain, diffraction (dispersion), and filtering. Therefore, this equation has numerously arisen in simulatations of optical systems \cite{Boardman_Chapter_2005,Boardman_2006,Grelu_NP_2012}. Moreover, it admits vast variety of sophisticated solutions in the form of stable localized structures representing the dissipative solitons \cite{Akhmediev_Chapter_2005}. Indeed, the solutions to the one-dimensional CGLE correspond to solitons with stationary \cite{Fauve_PRL_1990, van_Saarloos_PD_1992, Afanasjev_PRE_1996, Renninger_PRA_2008}, periodically, quasi-periodically, and aperiodically (chaotically) pulsating waveforms \cite{Deissler_PRL_1994, Soto-Crespo_PRL_2000, Akhmediev_PRE_2001}, exploding solitons \cite{Soto-Crespo_PRL_2000, Akhmediev_PRE_2001, Soto-Crespo_PLA_2001, Cundiff_PRL_2002, Descalzi_PRE_2011}, and solitons with periodical and chaotic spikes of extreme amplitude and short duration \cite{Chang_OL_2015, Chang_JOSAB_2015, Soto-Crespo_JOSAB_2017}. It also admits the existance of multisoliton solutions \cite{Akhmediev_PRL_1997} and dissipative solitons in a form of stable dynamic bound states \cite{Turaev_PRE_2007}. These different solutions coexist to each other when the parameters of equation belong to certain regions \cite{Afanasjev_PRE_1996, Soto-Crespo_PRL_2000, Akhmediev_PRE_2001,Soto-Crespo_PLA_2001,Descalzi_PhysicaA_2006}. Moreover, the basic CGLE model can easily be modified to account for such higher-order effects as fourth-order spectral filtering, third-order dispersion, and stimulated Raman scattering \cite{Soto-Crespo_PRE_2002, Achilleos_PRE_2016, Sakaguchi_OL_2018, Uzunov_PRE_2018} as well as to investigate the turbulent-like intensity and polarization rogue waves in a Raman fiber laser \cite{Sugavanam_LPR_2015}, stationary solitary pulses in a dual-core fiber laser \cite{Malomed_C_2007}, and the interaction of stationary, oscillatory and exploding counter-propagating dissipative solitons \cite{Descalzi_EPJ_2015,Descalzi_C_2018}.

The incorporation of a potential term into the CGLE has opened up new insights into the getting control over soliton waveforms. In particular, the diffusion-induced turbulence in systems near a supercritical Hopf bifurcation is simulated on the base of the CGLE with an additional term accounting for the global delayed feedback \cite{Battogtokh_PD_1996} and a gradient force \cite{Xiao_PRL_1998}. The comprehensive ideas of soliton management in Bose-Einstein condensates are discussed in \cite{Malomed_Book}, while the computational aspects of the related equations are presented in \cite{Caliari_CPC_2013,Antoine_CPC_2013,Antoine_CPC_2014,Antoine_CPC_2015}. In optics, the splitting of spatial solitons into two and more beams due to their scattering on a longitudinal defect \cite{Fratalocchi_PRE_2006}, an external delta potential \cite{Holmer_JNS_2007}, and a longitudinal potential barrier \cite{Yang_OE_2008} has been modelled in the framework of the the nonlinear Schr\"odinger equation with a control term. A sharp potential barrier \cite{He_JOSAB_2010}, umbrella-shaped \cite{Yin_JOSAB_2011}, and radial-azimuthal \cite{Liu_OE_2013} potentials have been added to the one- and two-dimensional cubic-quintic CGLE to study various scenarios of the dynamics of dissipative solitons in active bulk media with spatially modulated refractive indexes.

The account for an external magnetic field in nonlinear magneto-optic waveguides with different configurations gives one more prominent example of optical solitons governed by evolutionary equations with potential terms \cite{Boardman_1995,Boardman_1997,Boardman_2001,Boardman_2003,Boardman_2005,Boardman_2010}. In fact, the applied spatially inhomogeneous magnetic field locally breaks the time reversal symmetry leading to significantly different propagation conditions of counter-propagating light beams in such waveguides \cite{Boardman_Chapter_2005, Boardman_2006}. Further, this control approach over soliton waveforms has been developed to perform a  selective lateral shift within a group of stable noninteracting fundamental dissipative solitons \cite{OptLett_2017}, to replicate them \cite{PRE_2017}, and to induce the waveform transitions \cite{Chaos_2018}.

Since waveforms of stationary dissipative solitons show significant stability upon distortion effects they have been recognized as promising information carriers for new optical networks. However, the development of such systems requires getting the full control over the solitons waveforms. In this paper we further develop a mechanism to control waveforms of dissipative solitons. In particular, we employ the one-dimensional cubic-quintic CGLE with a potential term to perform the mutually-invertible waveform transitions between different stationary and moving dissipative solitons. These transitions are controlled by the potentials locally applied along the propagation distance. The used model appears in the theory of a planar magneto-optic waveguide in the Voigt configuration \cite{Boardman_Chapter_2005,Boardman_2006}.

The rest of the paper is structured as follows. In Section~\ref{S:2} we introduce the basic evolutionary equation with a potential term governing the soliton dynamics, where the potential controls the waveform transitions. In Section~\ref{S:3} we simulate mutual transitions between three different solitons coexisted in the system simultaneously. In particular, we perform the mutually-invertable transitions between so-called plain, composite, and moving pulses and find typical transverse spatial distributions of potentials with finite supporters along the propagation distance. Conclusions and final remarks are made in Section~\ref{S:4}.

\section{The Model of Controllable Waveform Transitions}
\label{S:2}
Having assumed the context of planar magneto-optic waveguides  \cite{Boardman_Chapter_2005, Boardman_2006, OptLett_2017} we write the one-dimensional cubic-quintic CGLE with a potential term as follows
\begin{align}
\label{CQCGLE}
\mathrm{i}&\frac{\partial\Psi}{\partial z}+\mathrm{i}\delta\Psi+\left(\frac{1}{2}-\mathrm{i}\beta\right)\frac{\partial^2\Psi}{\partial x^2} \nonumber \\ &+\left(1-\mathrm{i}\varepsilon\right)\left|\Psi\right|^2\Psi -\left(\nu-\mathrm{i}\mu\right)\left|\Psi\right|^4\Psi + Q(x,z)\Psi = 0,
\end{align}
where $\Psi\left(x,z\right)$ is the complex slowly varying soliton envelop of the transverse $x$ and longitudinal $z$ coordinates. All coefficients in Eq.~\eqref{CQCGLE} are assumed to be positive. Therefore, $\delta$ and $\beta$ stand respectively for the linear absorption and diffusion, $\nu$ accounts for the self-defocusing effect due to the quintic nonlinearity, while $\varepsilon$ and $\mu$ are the cubic gain and quintic loss coefficients, respectively.

The potential $Q(x,z)$ in the last term of Eq.~\eqref{CQCGLE} describes an arbitrary external linear control whose particular spatial distribution can be adjusted depending on the physical origin of modelled system. In this regard, the optical applications can be mentioned, where the potential $Q(x,z)$ accounts for the linear magneto-optic effect in waveguides \cite{Boardman_Chapter_2005, Boardman_2006, OptLett_2017} and spatial modulation of the refractive index \cite{He_JOSAB_2010, Yin_JOSAB_2011, Liu_OE_2013, Liu_OL_2010, Liu_OE_2011}. Independently on particular application, it is logically to assume that an external control influence acts locally along the propagation distance, e.i. it has a finite supporter along the $z$ axis. Therefore, for simplicity  the longitudinal dependence of potential is chosen in the form of a piecewise constant function
\begin{equation}
\label{Q}
Q(x,z) = q(x)\left[h(z-z_1)-h(z-z_2)\right],
\end{equation}
where $q(x)$ defines the dependence of potential on the transverse coordinate $x$, $h(\cdot)$ is the Heaviside step function, and $z_1<z_2$ are two points on the $z$ axis at which the potential is respectively switched on and off. Particular transverse distributions $q(x)$ are specified and discussed later on.

In order to solve Eq.~\eqref{CQCGLE} we use exponential time differencing numerical schemes as well as their Runge-Kutta modifications with second- and fourth-order accuracy in the Fourier domain \cite{Cox_2002}. The Fourier transform is applied with respect to the transverse coordinate $x$ such that the complex amplitude $\Psi(x,z)$ corresponds to its Fourier spectrum $\hat{\Psi}(k_x,z)$. Having applied the fast Fourier transform we implicitly impose the following periodic boundary condition 
\begin{equation}
\label{BC}
\Psi(x,z) = \Psi(x+L_x,z),~~~~~~~~\forall(x,z)\in\mathbb{R}\times[0,+\infty),
\end{equation}
for some $L_x>0$. Thus, the computational domain is bounded by the finite rectangular $[-L_x/2,L_x/2]\times[0,L_z]$, where the width $L_x=140$ is chosen to ensure that all non-negligible parts of waveforms belong to the domain, while the length is typically set as $L_z=300$. The computational domain is sampled with $N_x=2^{12}$ points along the $x$ axis, while the step $\Delta z=10^{-3}$ is taken to discretize it along the propagation distance.

\begin{figure}[htbp]
\centerline{\includegraphics[width=0.6\linewidth]{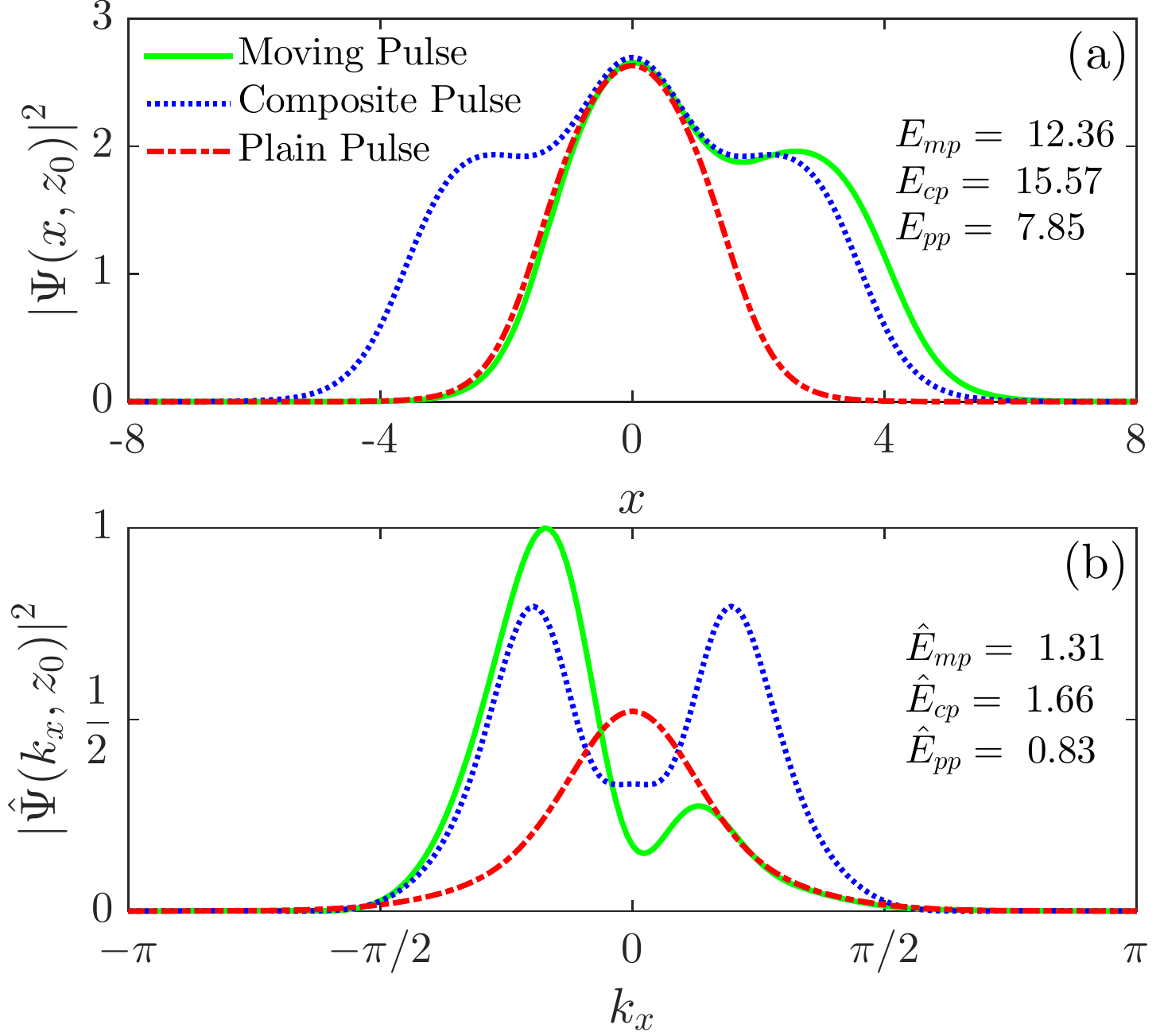}}
\caption{Three forms of stable dissipative solitons coexisted for the same set of equation coefficients: (a) intensities; (b) normalized power spectra.}
\label{fig1}
\end{figure} 

All numerical simulations have been carried out using the same fixed values of the coefficients of Eq.~\eqref{CQCGLE}. Namely, they are $\beta=0.5$, $\delta=0.1$, $\mu=0.75$, $\nu=0.1$, and $\varepsilon=1.85$. This set of values allows the coexistence of three different forms of stable dissipative solitons \cite{Afanasjev_PRE_1996}. Two of them are stationary solutions to Eq.~\eqref{CQCGLE} with zero potential, which are called \textit{plain} and \textit{composite} pulses, while the last one is a uniformly translating along the transverse direction \textit{moving} pulse \cite{Afanasjev_PRE_1996}. The plain pulse is also known as the fundamental soliton. In other words, the waveforms of all these pulses do not change their profiles when pulses propagate along the $z$ axis, however moving pulse has additional transverse drift, while the plain and composite pulses are unmovable along the $x$ axis. The intensity distributions and normalized power spectra of these three pulses are plotted in Fig.~\ref{fig1}, where the dash-dot red, dotted blue, and solid green lines indicate the plain, composite, and moving pulses, respectively. Both intensities in Fig.~\ref{fig1}(a) and spectra in Fig.~\ref{fig1}(b) are calculated at certain point $z_0$ on the $z$ axis when no applied potential, i.e. $Q(x,z)=0$. The intensity profiles of the plain and composite pulses are unchangeable and they have the same transverse location at any point on the $z$ axis as long as the equation coefficients are fixed and the potential is switched off. The intensity profile of the moving pulse is also unchangeable under the fixed parameters, however it uniformly moves along the negative direction of the $x$ axis. Thus, Fig.~\ref{fig1}(a) shows the instantaneous location of the moving pulse. It is clear that the pulse spectra are unchangeable and unmovable in the Fourier domain. Additionally, Fig.~\ref{fig1} contains the energies of all pulses calculated both in the coordinate space ($E_{pp}$, $E_{cp}$, $E_{mp}$) and the Fourier domain ($\hat{E}_{pp}$, $\hat{E}_{cp}$, $\hat{E}_{mp}$), where the subindexes stand for the abbreviations of the pulses.

In fact, there always exists a pair of moving pulses, whose waveforms drift along the $x$ axis in opposite directions. For the brevity, we only consider the moving pulse with the drift along the negative direction of the $x$ axis because the other case is straightforward - everything is symmetric with respect to the plane $x=0$.   

\section{Mutual Waveform Transitions}
\label{S:3}
Looking at the plain, composite and moving pulses from the viewpoint of the theory of dynamical systems one can consider these coexisted solutions as different attractors (stable fixed points) in an infinite-dimensional phase space of the system~\eqref{CQCGLE} \cite{Akhmediev_Chapter_2005}. Indeed, each of the pulses can equally well be excited using any initial conditions, which start phase trajectories in vicinities of those fixed points. For example, in order to excite the composite pulse one can use the following real function
\begin{equation} \label{IC}
\Psi_0(x)=1.5\exp\left(-\frac{x^2}{9}\right).
\end{equation}

As long as all parameters of the system are fixed a phase trajectory cannot change its basin of attraction. However, switching on the potential we can perturb a soliton waveform in such a way that the perturbed waveform corresponds to a new phase trajectory in a vicinity of another attractor. Then switching off the applied potential we remove the influence induced by the potential and the waveform quickly evolve to a new unperturbed state. In general, this idea of controllable waveform transitions between different coexisted dissipative solitons have been considered in \cite{Chaos_2018}. In this section we compliment the results of \cite{Chaos_2018}. Namely, we seek the transverse distributions $q(x)$ in~\eqref{Q} to perform all possible waveform transitions between plain, composite, and moving pulses, which are not sensitive to the moment at which the potential~\eqref{Q} is switched off. In other words we discuss a regular waveform transition whose outcome does not depend on the point $z_2$ if the longitudinal length of the potential $z_2-z_1$ exceeds some critical value. To be specific, we always assume in~\eqref{Q} that $z_1=100$ and $z_2=200$.

\subsection{Transitions to Plain Pulse}
\label{S:31}
First of all, we consider the induced waveform transitions of the composite and moving pulses whose waveforms are relatively complex to the plain pulse with the simplest waveform. One can see that the plain pulse has a symmetric bell-shaped waveform. Therefore, a potential appropriated for these transitions should squeeze the waveforms of both composite and moving pulses to such a profile along the $x$ axis. Thus, it is logically to choose the potential spatial distribution $q(x)$ in the form of one-peaked symmetric function. Accounting for the physical reasons \cite{Boardman_1997}, we take the $\mathrm{sech(x)}$ as a trial function and write down the inhomogeneous transverse distribution of the potential~\eqref{Q} as follows 
\begin{equation}
\label{q1s}
q_{1\textrm{s}}(x) = A~\mathrm{sech}\left(\frac{x-x_p}{x_w}\right),
\end{equation}
where $A$ and $x_w$ adjust the height and width of the potential, respectively, while $x_p$ defines the transverse position of the peak.

\begin{figure}[htb]
\centerline{\includegraphics[width=1\linewidth]{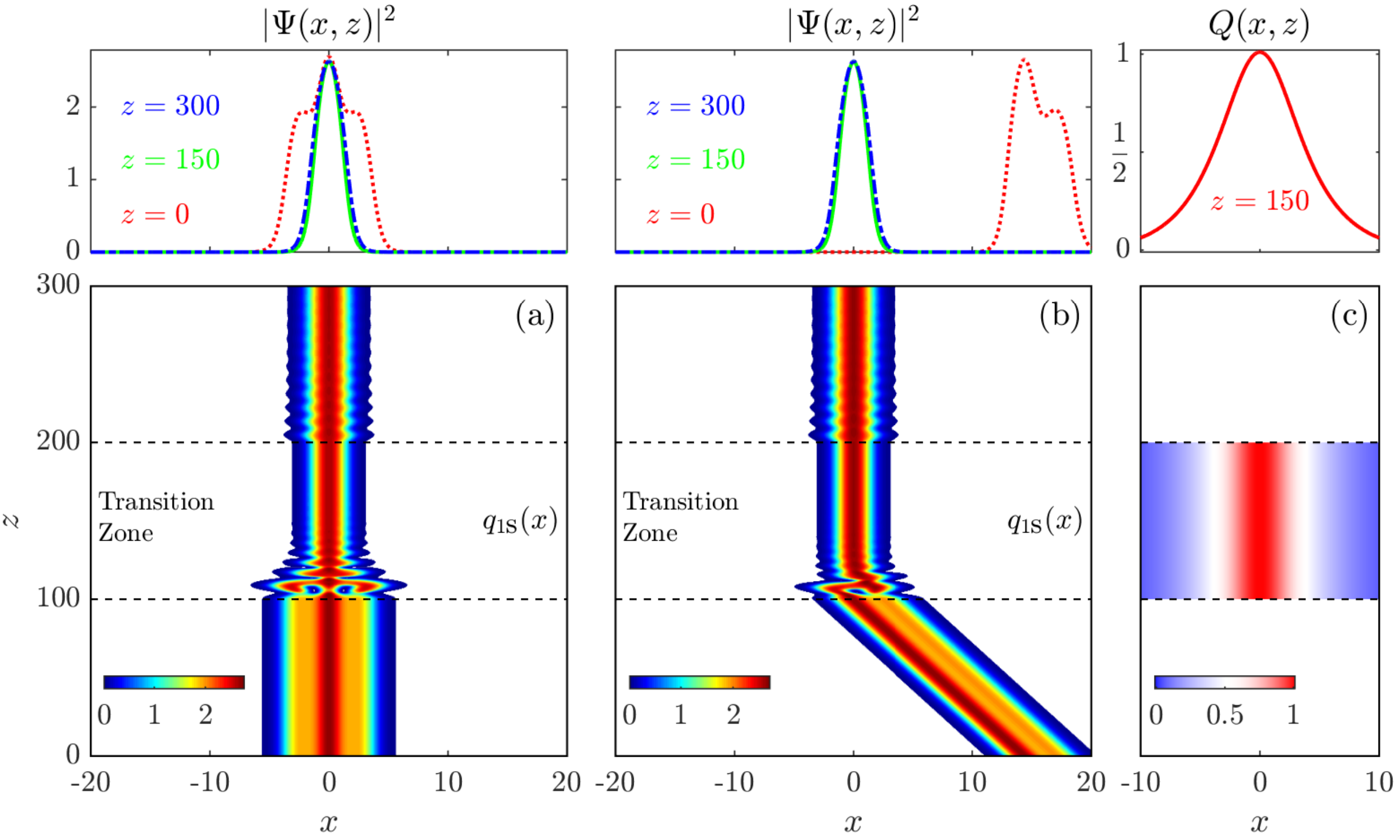}}
\caption{Regular waveform transitions of the composite (a) and moving (b) pulses to the plain pulse induced by the potential (c) with one-peaked symmetric function $q_{\textrm{1s}}(x)$.}
\label{fig2}
\end{figure}

For a wide range of the parameters, the potential~\eqref{Q} with the function \eqref{q1s} has a typical profile to transit both the composite and moving pulses to the plain pulse. In particular, these transitions are presented in Fig.~\ref{fig2}, where the parameters of~\eqref{q1s} are set as $A=1$, $x_w=3$, and $x_p=0$. Precisely, Fig.~\ref{fig2}(a) and Fig.~\ref{fig2}(b) show the two-dimensional intensity plots of complex envelops $|\Psi(x,z)|^2$, which respectively indicate the transitions of the composite and moving pulses to the plain pulse. The upper panels in Fig.~\ref{fig2}(a) and Fig.~\ref{fig2}(b) contain three cross sections of intensity plots, which demonstrate the instantaneous intensity distributions of solitons before ($z=0$), during ($z=150$), and after ($z=300$) transitions. Fig.~\ref{fig2}(c) shows two-dimensional plot of the potential $Q(x,z)$ applied to induce those transitions, while its upper panel contains the plot of the transverse function~\eqref{q1s}.

Both waveform transitions presented in Fig.~\ref{fig2} have been performed according to the same scenario. At first, some original soliton (composite or moving pulse) appears at the section $z=0$ and freely ($Q(x,z)=0$) propagates towards the section $z=z_1=100$ at which the transition zone begins because the potential \eqref{Q}, \eqref{q1s} starts its influence upon soliton waveforms at this section. Later, the potential with the properly chosen profile $q_{\textrm{1s}}(x)$ dramatically perturbs the input waveforms that they collapse to the same bell-shaped waveform irrespectively to the different inputs. Finally, at the section $z=z_2=200$ the potential is switched off and the transition zone is finished allowing to the perturbed waveforms freely evolve to the plain pulse. It is obvious that the potential \eqref{Q}, \eqref{q1s} transits the input plain pulse waveform to itself.

In both cases the transition zones can be separated into two stages. The first one is the relatively short transient stage, which contains a fast transition of the input waveform to a new forced steady-state waveform under the influence of the applied potential. It starts as soon as the potential is applied ($z=z_1=100$) and goes until the strong and fast pulsations of the perturbed waveform have decayed ($z\approx 140$). The second stage is the steady-state stage, where the forced steady-state waveform gradually approaches to its limit stationary state. This stage comes to replace the transient stage and goes as long as the potential is applied.

In general, if the potential is switched off somewhere within the transient stage then the outcome of such transition may strongly depend on the section $z_2$ at which the potential was switched off. In contrast, switching off the potential within the steady-state stage we always obtain the same outcome for the given waveform transition. Such transitions, which are stable with respect to small changes in the potential profile are known as regular waveform transitions \cite{Chaos_2018}.

Using the language of dynamical systems we can say that the influence of the potential upon the soliton waveforms is used as an external local control capable to move the different phase trajectories (they correspond to input waveforms) within a phase space of the system \eqref{CQCGLE} with zero potential. In Fig.~\ref{fig2}, the potential profile (external control) is chosen to move all such input trajectories from their original basins of attractions to a basin of another attractor that corresponds to the plain pulse. This control mechanism is also used to transit different input waveforms to the output waveforms in the form of composite or moving pulses as discussed below.

\subsection{Transitions to Composite Pulse}
\label{S:32}
Now we consider features of the transverse potential function $q(x)$ suitable to transit the waveforms of both the plain and moving pulses to the composite pulse. Looking at Fig.~\ref{fig1} we observe that the intensity plot and power spectrum of the composite pulse are symmetric. Moreover, its spectrum has two distinct peaks, while its waveform consists of two symmetrical fronts and a hill between them. Therefore, the sought transverse potential function $q(x)$ should symmetrically diverge the input wavewofms into two fronts along the $x$ axis. Due to these reasons we seek the potential function $q(x)$ in the form of two-peaked symmetric function. Combining two $\mathrm{sech}(x)$ functions we get
\begin{equation}
\label{q2s}
q_{2\textrm{s}}(x)=A~\mathrm{sech}\left(\frac{x+x_p}{x_w}\right)+A~\mathrm{sech}\left(\frac{x-x_p}{x_w}\right).
\end{equation}

\begin{figure}[htb]
\centerline{\includegraphics[width=1\linewidth]{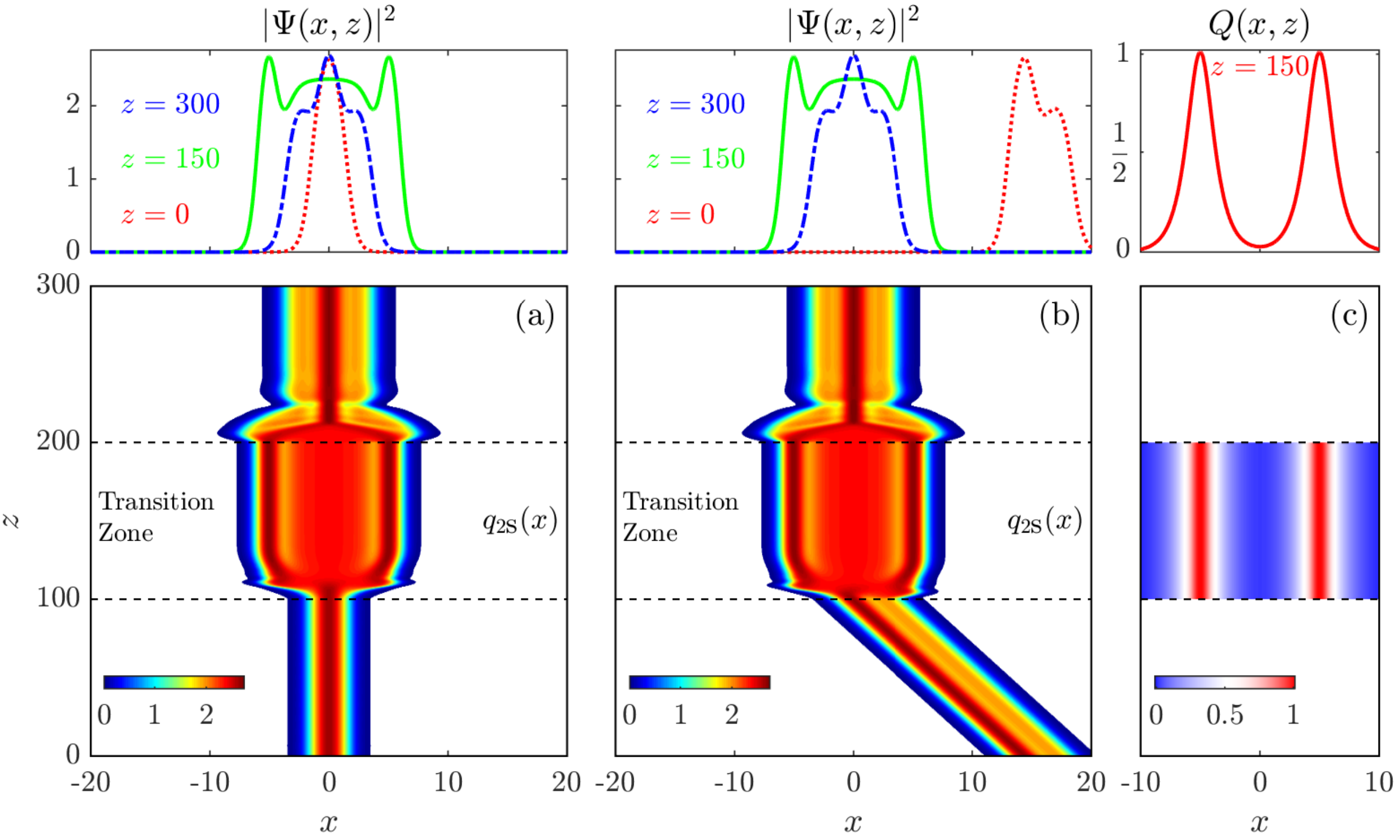}}
\caption{Regular waveform transitions of the plain (a) and moving (b) pulses to the composite pulse induced by the potential (c) with two-peaked symmetric function $q_{\textrm{2s}}(x)$.}
\label{fig3}
\end{figure}

For a reasonable set of parameters, the potential \eqref{Q}, \eqref{q2s} possesses all necessary features to perform waveform transitions of the plain and moving pulses to the composite pulse. Indeed, setting up $A=1$, $x_w=1$, and $x_p=5$ we perform these transitions as presented in Fig.~\ref{fig3}. The identical transition of the composite pulse to itself can also be induced by the potential \eqref{Q}, \eqref{q2s} but it is skipped due to its triviality. All these transitions have been performed according to the same scenario, which is described in Section~\ref{S:31}. 

The plots in Fig.~\ref{fig3} are similar to those presented in Fig.~\ref{fig2} excepting of now we use other initial waveforms and apply the potential with another transverse distribution $q_{\textrm{2s}}(x)$ to move the input trajectories from their original basins of attraction to a vicinity of the composite pulse attractor. 

\subsection{Transitions to Moving Pulse}
Finally, we consider the waveform transitions of the stationary pulses to the moving pulse and discuss the properties of the transverse potential function $q(x)$ suitable to induce these transitions. Comparing the instantaneous waveforms and spectra depicted in Fig.~\ref{fig1} we can roughly conclude that the moving pulse is a kind of asymmetric version of the composite pulse. Thus, the sought transverse potential function should have two-peaked asymmetric profile. Using a combination of two $\mathrm{sech}(x)$ functions we write down this function as follows
\begin{equation}
\label{qas}
q_{\textrm{as}}(x) = A~\mathrm{sech}\left(\frac{x+x_p}{x_w}\right)+B~\mathrm{sech}\left(\frac{x-x_p}{x_w}\right).
\end{equation}

\begin{figure}[htb]
\centerline{\includegraphics[width=1\linewidth]{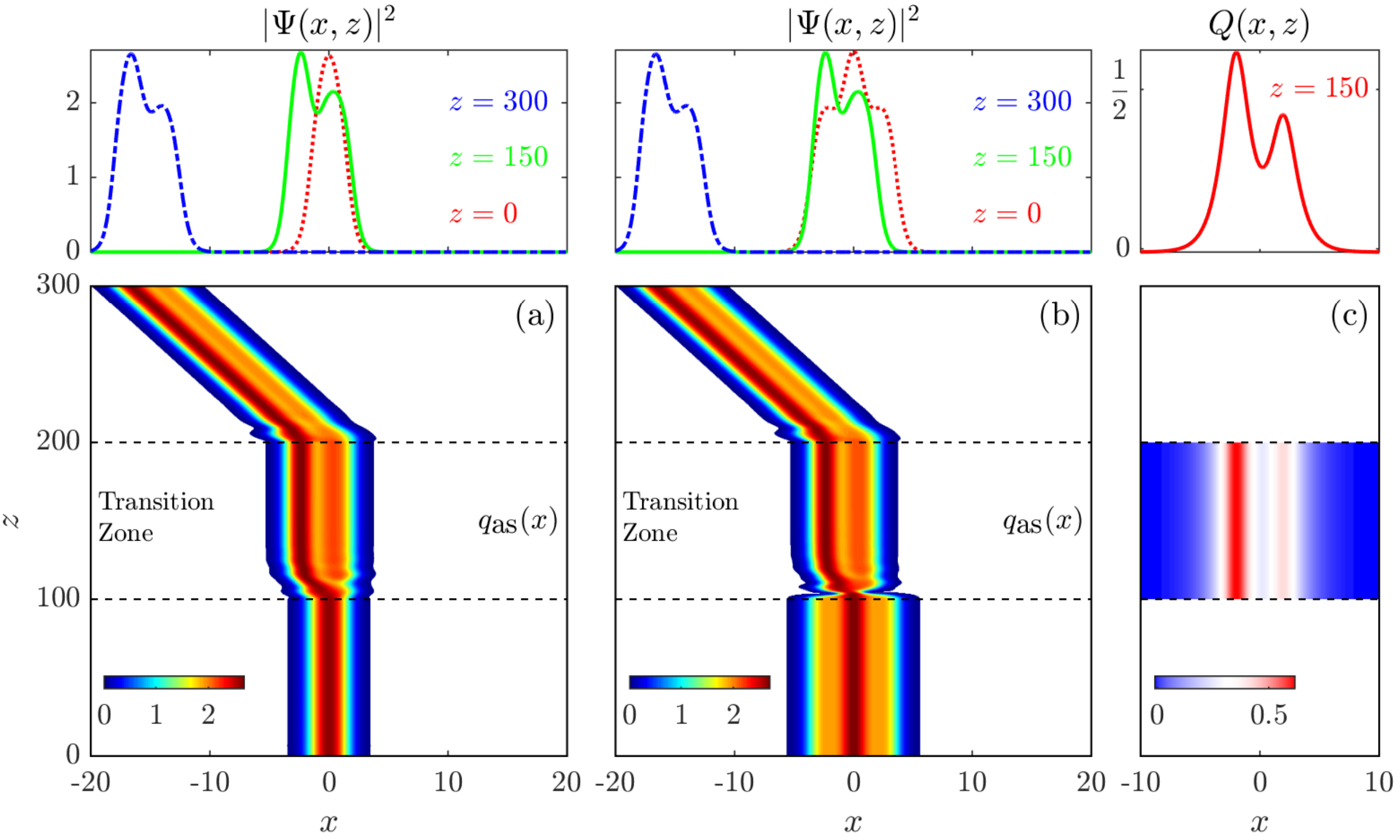}}
\caption{Regular waveform transitions of the plain (a) and composite (b) pulses to the moving pulse induced by the potential (c) with two-peaked asymmetric function $q_{\textrm{as}}(x)$.}
\label{fig4}
\end{figure}

Having substituted the numerical values $A=0.6$, $B=0.4$, $x_w=1$, and $x_p=2$ into \eqref{qas}
we apply the potential \eqref{Q}, \eqref{qas} to perform the waveform transitions of the plain and composite pulses to the moving one, as presented in Fig.~\ref{fig4}. This potential also perform the identical transition of the moving pulse to itself, which is trivial. In order to perform all these transitions we again reiterate the same scenario but with another transverse potential function $q_{\textrm{as}}(x)$, which is suitable to induce the later transitions.

The numerical simulations presented in Fig.~\ref{fig4} demonstrate the transitions of two stationary pulses to the moving pulse with uniform translating along the negative direction of the $x$ axis. This drift direction is defined by the pulse asymmetry - the pulse front is higher than its tail. In our simulations we set the left peak of the potential function $q_{\textrm{as}}(x)$ greater than the right one ($A>B$) that lead to the outcomes presented in Fig.~\ref{fig4}. However, if we exchange the values of the peak heights and simulate again these transitions excepting of now $A<B$ then we would have the outcomes in the form of the moving pulse with uniform translating along the positive direction of the $x$ axis. Thus, we conclude that a higher left (right) peak in the potential function \eqref{qas} leads to the moving pulse with left (right) drift along the $x$ axis, while the symmetric profile leads to the stationary pulse.

\section{Conclusions}
\label{S:4}
In series of numerical simulations presented in Figs.~\ref{fig2} - \ref{fig4} we have demonstrated all possible combinations of  mutually-invertable waveform transitions between the plain, composite and moving pulses. In each of our simulations we applied the potential~\eqref{Q} with typical transverse potential function $q(x)$, which is taken in the form of scaled $\mathrm{sec}(x)$ functions. These particular profiles \eqref{q1s}-\eqref{qas} have been chosen due to reasons arisen in optical applications \cite{Boardman_1997}. However, they contain all the important features of the potential profiles to be suitable for inducing the waveform transitions.

\newcommand{\specialcell}[2][c]{\begin{tabular} [#1]{@{}c@{}}#2 \end{tabular}}
\begin{table}[htbp]
\label{tab1}
\begin{tabular}{|c||*{3}{c|}}
\hline
\diagbox{Input}{Output} & Plain Pulse & Composite Pulse & Moving Pulse \\
\hline
\hline
\specialcell{Plain Pulse\\ \\ Composite Pulse \\ \\ Moving Pulse} & \specialcell{One-peaked\\symmetric\\potential\\\begin{minipage}{0.2\textwidth}
      \includegraphics[width=\linewidth, height=20mm]{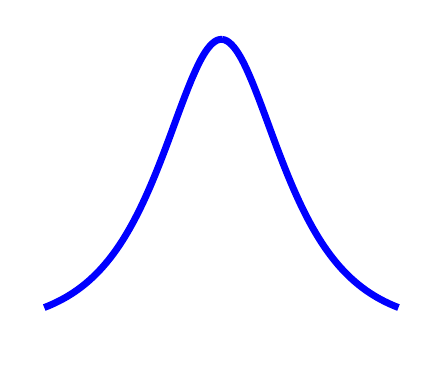}
\end{minipage}} &
\specialcell{Two-peaked\\symmetric\\potential\\\begin{minipage}{0.2\textwidth}
      \includegraphics[width=\linewidth, height=20mm]{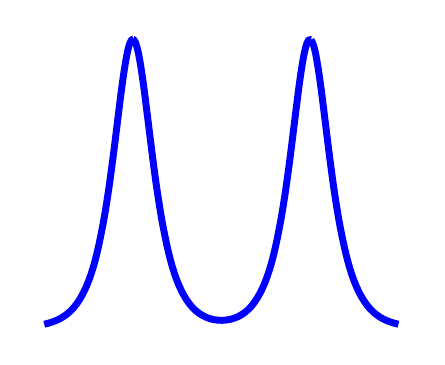}
\end{minipage}} &
\specialcell{Two-peaked\\asymmetric\\potential\\\begin{minipage}{0.2\textwidth}
      \includegraphics[width=\linewidth, height=20mm]{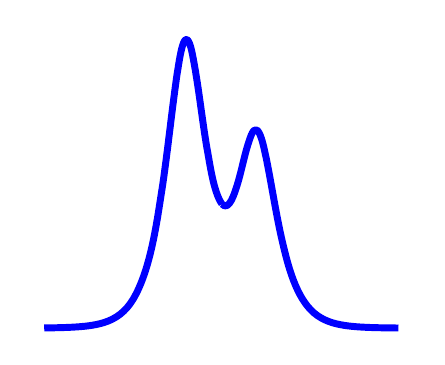}
\end{minipage}} \\
\hline
\end{tabular}
\caption{Typical transverse distributions of the potentials for inducing the waveform transitions between the plain, composite, and moving pulses.}
\end{table}

Taking into account our numerous simulations with different transverse potential functions $q(x)$ whose heights and widths are comparable with dimensionless sizes of propagating solitons we conclude that:
\begin{itemize}
\item the one-peaked symmetric potential is appropriate to transit the plain, composite and moving pulses to the plain pulse;
\item the two-peaked symmetric potential is appropriate to transit the plain, composite and moving pulses to the composite pulse;
\item the two-peaked asymmetric potential is appropriate to transit the plain, composite and moving pulses to the moving pulse.
\end{itemize}

Schematically these conclusions are summarized in Table~1.

The presented results on mutual transitions can be useful for the development of such new optical components and devices as beam splitters, demultiplexers, logic gates, etc., where the dissipative optical solitons are employed as information carriers.


$$~$$
\textbf{Acknowledgements}.

The author is grateful for a support from Jilin University and useful discussions with Prof. V.~Tuz.

\bibliographystyle{model1-num-names}
\bibliography{Soliton}

\end{document}